\documentclass[aps,twocolumn,a4paper,floatfix,tightenlines,superscriptaddress,pra,longbibliography]{revtex4-1}
\usepackage[toc,page]{appendix}
\usepackage{braket}
\usepackage{epsfig,graphicx,times}
\usepackage{amstext}
\usepackage{amsmath}
\usepackage{amssymb}
\usepackage{mathrsfs}
\usepackage{dcolumn}
\usepackage{bm}
\usepackage[tight]{subfigure}
\usepackage[colorlinks,linkcolor=red,anchorcolor=blue,citecolor=blue,urlcolor=red]{hyperref}
\usepackage{color}
\usepackage{siunitx}
\usepackage{appendix}
\usepackage{placeins}
\usepackage{textcomp}
\usepackage{bbold}
\usepackage{float}
\usepackage{verbatim}
\usepackage{minitoc}
\usepackage[dvipsnames]{xcolor}

\usepackage{soul}
\usepackage[framemethod=tikz]{mdframed}
\usepackage{CJKutf8}

\begin{document}
\begin{CJK}{UTF8}{<font>}

\title{Witnessing entanglement via the geometric phase in a impurity-doped Bose-Einstein condensate}

\author{Xia Wu}
\affiliation{Key Laboratory of Low-Dimensional Quantum Structures and Quantum Control of Ministry of Education, Department of Physics and Synergetic Innovation Center for Quantum Effects and Applications, Hunan Normal University, Changsha 410081, China}

\author{Shao-Peng Jia}
\affiliation{Key Laboratory of Low-Dimensional Quantum Structures and Quantum Control of Ministry of Education, Department of Physics and Synergetic Innovation Center for Quantum Effects and Applications, Hunan Normal University, Changsha 410081, China}

\author{Cui-Lu Zhai}
\affiliation{Key Laboratory of Low-Dimensional Quantum Structures and Quantum Control of Ministry of Education, Department of Physics and Synergetic Innovation Center for Quantum Effects and Applications, Hunan Normal University, Changsha 410081, China}

\author{Le-Man Kuang}\email{lmkuang@hunnu.edu.cn}
\affiliation{Key Laboratory of Low-Dimensional Quantum Structures and Quantum Control of Ministry of Education, Department of Physics and Synergetic Innovation Center for Quantum Effects and Applications, Hunan Normal University, Changsha 410081, China}

\begin{abstract}

We propose a theoretical scheme to witness quantum entanglement via the geometric phase  in an impurity-doped Bose-Einstein condensate (BEC), which is a micro-macro quantum system  consisting of two Rydberg impurity qubits and the BEC. We calculate the geometric phase of the impurity qubits in the presence of the initial micro-micro  and micro-macro entanglement, respectively. It is demonstrated that the geometric phase of the impurity qubits can witness not only inter-qubit micro-micro entanglement, but also  qubit-BEC micro-macro entanglement. Our work provide a new insight to witness micro-micro  and micro-macro entanglement in a impurity-doped BEC.

\end{abstract}

\maketitle

\section{Introduction}

Quantum entanglement lies at the heart of quantum physics and  the  emerging second quantum revolution \cite{Dowling,Deutsch}, and is a key resource for quantum information science and technology \cite{Chitambar,Horodecki}.  Its detection is thus of crucial importance and has been studied extensively, notably with so-called entanglement witnesses \cite{Horodecki}. The fact that there exist entanglement witness for every entangled state \cite{Horodecki1} has raised the importance on a theoretical point of view even further \cite{Chruciski,Hyllus}.
In particular, it is one of fascinating problems in the field of quantum physics and quantum information  to detect  quantum entanglement of a micro-macro system \cite{1,JQ2009,2,3,4,5,6,7,JQ2015,JQ2016,JQL2016,JQ2012,8,9}.
The difficulties inherent in such a question are manifolds, and they are related  not only to quantum decoherence induced by the surrounding environment \cite{10,FD2008,NS2011,YJ2019,11,QS2017,JB2017,JBY2017,12,LM2007,13}, but also to a measurement precision sufficient to observe quantum effects at such macroscales.

A Bose-Einstein condensate (BEC) doped with  impurities  \cite{Ng,Balewski,Schmidt,Johnson,LiQIP,Yuan1,Song}  provides an ideal platform for the  study of  micro-micro and micro-macro entanglement where microscopic impurities meets a macroscopic matter, the BEC. As the interaction among Rydberg impurity atoms \cite{23,24,25,26,27,Wang} can be tailored by electric fields and microwave fields  \cite{28,29} while the BEC allows for an extremely precise control of interatomic interactions by manipulating $s$-wave scattering length \cite{30,31}, they can build a precisely controllable micro-macro quantum systems  \cite{32}.

Since Berry demonstrated that a geometric phase can be created in a quantum system undergoing a cyclic adiabatic evolution \cite{Berry}, much attention has been paid to the geometric phase \cite{Aharonov,Anandan,Samuel}. Geometric phases of pure states have been generalized to  mixed states \cite{Uhlmann,Pati,Singh} and open quantum systems \cite{Tong}. Important  experimental progresses have been made on the geometric phase \cite{Leek,Carollo,Ericsson,Marzlin,Kamleitner,Du,Ericsson2,Whitney,Bassi,Yi,Rezakhani,Lombardo,Sarandy,Goto,Moller,Buri}.
In particular, the rise of quantum information science has opened up a new direction for applications of the geometric phase  as well as triggered new insights into its physical nature and applications, such as geometric  quantum computation \cite{Zanardi,Pachos,Jones,Zu} and a new type of topological phases for entangled quantum systems \cite{Milman,Oxman}.

In the present paper, we want to propose a theoretical scheme to witness micro-micro and micro-macro entanglement via the geometric phase of the impurity qubits in an impurity-doped BEC, which consists of the BEC and two Rydberg impurity atoms treated as qubits. We calculate the geometric phase of the impurity qubits and obtain direct relations between  the geometric phase and quantum entanglement. It is shown that the geometric phase  of the impurity qubits can witnees  not only the micro-micro entanglement between two impurity qubits, but also micro-macro entanglement between the impurity qubits and the BEC.

The remainder of this paper is organized as follows. In Sec. \uppercase\expandafter{\romannumeral2}, we introduce the impurities-doped BEC model consisting of the BEC and two Rydberg impurities. We present an analytical solution of the impurities-doped BEC model for a general initial state of  the model. In Sec. \uppercase\expandafter{\romannumeral3}, we calculate  the geometric phase of the impurity qubits when the two qubits are initially in an entangled state while qubits and the BEC  are initially unentangled. It is indicated that the geometric phase can witness micro-micro entanglement between two Rydberg impurities.   In Sec. \uppercase\expandafter{\romannumeral4}, we investigate  the geometric phase of the impurity qubits when the qubits and the BEC  are initially  in entangled states. We find a direct relation between micro-macro entanglement and the geometric phase of the impurity qubits, and  demonstrate that the geometric phase can witness micro-micro entanglement. Finally, Sec.\uppercase\expandafter{\romannumeral5} is devoted to some concluding remarks.

\section{\label{Sec:2}The impurity-doped BEC Model}

The impurity-doped BEC system under our consideration consists  of a BEC and  two localized Rydberg impurity atoms  immersed in the BEC \cite{Zli}. The two separated Rydberg impurities  are frozen in place and they interact with each other via a repulsive van der waals interaction \cite{28,29}. The relevant internal level structure for each Rydberg atom is given by the atomic ground state $|0\rangle$ and the excited Rydberg state $|1\rangle$, which form an effective two-level system, i.e., an impurity qubit. The Hamiltonian of two Rydberg impurities in the absence of the external laser field  \cite{29} is given by
\begin{eqnarray}
H_R=\frac{\omega}{2}(\sigma_z^1+\sigma_z^2)+J\sigma_z^1\sigma_z^2,
\end{eqnarray}
where $\omega$ is the transition frequency between two internal states of each Rydberg impurity atom, the second term accounts for the van der Waals  interaction between the Rydberg impurities with the coupling strength $J=C_6/R^6$ where $R$ is the distance between two  localized Rydberg impurities, $C_6\propto \bar{n}^{11}$ with $\bar{n}$ being the principal quantum number of the Rydberg excitation.  We have set $\hbar=1$ in Hamiltonian (1) and through out the paper.

\begin{figure}[htp]
\includegraphics[width=6.2cm,height=6.0cm]{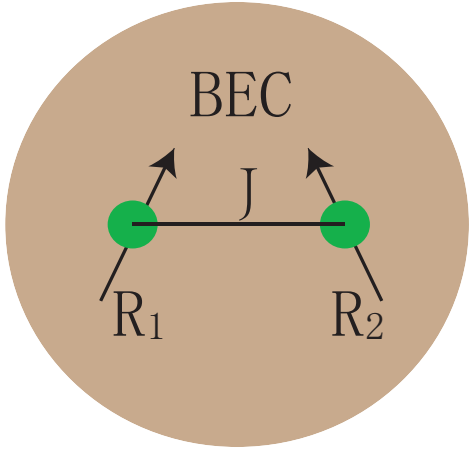}
\caption{Schematic diagram of two Rydberg impurities immersed in a Beose-Einstein condensate. Here $J$ denotes the van der Waals  interaction between  two Rydberg impurities.}
\end{figure}

Under the single-mode approximation, the Hamiltonian of a BEC confined in a trapping potential  has a Kerr-interaction form  given by
\begin{eqnarray}
H_B=\omega_b a^\dag a+\chi a^{\dag}a^{\dag} aa,
\end{eqnarray}
where $\omega_b$ is the mode frequency of the BEC which depends on the BEC mode function and the trapping potential, the nonlinear coupling constant  describes the inter-atomic $s$-wave scattering interaction in the BEC.

The two Rydberg impurities interact with the BEC via coherent collisions. The impurity-BEC interaction Hamiltonian can be given by
\begin{eqnarray}
H_I=\frac{\lambda}{2}(\sigma_z^1+\sigma_z^2)a^\dag a,
\end{eqnarray}
and $\lambda$ is the interaction strength.

Hence, The Hamiltonian of the total system including the two  Rydberg impurities  and the BEC is given by
\begin{align}
H=&H_R + H_B +H_I,
\end{align}
which is a diagonal Hamiltonian with the following eigenvalues and eigenstates
\begin{eqnarray}
E_{ijn}&=&\frac{1}{2}\omega\left[(-1)^{i} + (-1)^{j}\right] +\omega_b n + (-1)^{i+j}J \nonumber \\
&&+\frac{1}{2}\lambda\left[(-1)^{i} + (-1)^{j}\right]n  + \chi n(n-1), \\
|\psi\rangle_{ijn}&=&|i j n\rangle,
\end{eqnarray}
where  $|i j n\rangle=|i\rangle\otimes|j\rangle\otimes|n\rangle$ with $|i\rangle (|j\rangle)$ is a eigenstate of $\sigma_z^i (\sigma_z^j)$ with $i=0,1 (j=0,1)$, and $|n\rangle$ is a Fock state with $n=0, 1, 2, \cdots, \infty$.

Assume that the two impurity qubits and the BEC are initially in the following state
\begin{eqnarray}
\left\vert \Psi (0)\right\rangle =(c_{0}\left\vert 00\right\rangle
+c_{1}\left\vert 11\right\rangle +c_{2}\left\vert 01\right\rangle
+c_{3}\left\vert 10\right\rangle )\otimes \left\vert \alpha \right\rangle
\qquad \newline,
\end{eqnarray}
where $|\alpha\rangle$ is the Glauber coherent state defined by
\begin{eqnarray}
\left\vert \alpha \right\rangle =e^{-\frac{\left\vert \alpha \right\vert ^{2}%
}{2}}\underset{n=0}{\overset%
{\infty }{\sum }}\frac{\alpha ^{n}}{\sqrt{n!}}%
\left\vert n\right\rangle \newline,
\end{eqnarray}
and the superposition coefficients satisfy the normalization condition
\begin{eqnarray}
\left\vert c_{0}\right\vert ^{2}+\left\vert c_{1}\right\vert ^{2}+\left\vert
c_{2}\right\vert ^{2}+\left\vert c_{3}\right\vert ^{2}=1\newline.
\end{eqnarray}

It is straightforward to obtain the wavefunction of the impurity-doped BEC at an arbitrary time with the following expression
\begin{eqnarray}
 \vert\Psi (t)\rangle &=&c_{0}\vert 00\rangle \otimes
\vert \varphi _{0}(t)\rangle +c_{1}\vert 11\rangle
\otimes \vert \varphi _{1}(t)\rangle \nonumber \\
&&+c_{2}\vert01\rangle \otimes \vert \varphi _{2}(t)\rangle
+c_{3}\vert 10\rangle \otimes \vert \varphi
_{3}(t)\rangle,
\end{eqnarray}
where $\vert \varphi _{i}(t)\rangle(i=0,1,2,3)$ are four generalized coherent states given by
\begin{eqnarray}
\left\vert \varphi _{i}(t)\right\rangle =e^{-\frac{\left\vert \alpha
\right\vert ^{2}}{2}}\underset{n=0}{\overset%
{\infty }{\sum }}e^{-it\theta
_{i}(n)}\frac{\alpha ^{n}}{\sqrt{n!}}\left\vert n\right\rangle \newline,
\end{eqnarray}
where the four running frequencies are given by
\begin{eqnarray}
\theta _{0}(n)&=&-\omega +J-(\lambda -\omega _{b})n+\chi n(n-1), \nonumber\\
\theta _{1}(n)&=&\omega +J+(\lambda +\omega _{b})n+\chi n(n-1),\nonumber\\
\theta _{2}(n)&=&\theta _{3}(n)=-J+\omega _{b}n+\chi n(n-1).
\end{eqnarray}

From Eq. (10) we can obtain the reduced density operator of the two impurity qubits with the following form
\begin{equation}
\rho(t)=\left(
  \begin{array}{cccc}
    \rho _{00}(t) & \rho _{01}(t) & \rho _{02}(t) & \rho _{03}(t)\\
    \rho _{10}(t)  & \rho _{11}(t)  & \rho _{12}(t) & \rho _{13}(t)\\
    \rho _{20}(t) & \rho _{21}(t) & \rho _{22}(t) & \rho _{23}(t) \\
    \rho _{30}(t) & \rho _{31}(t) & \rho _{32}(t) & \rho _{33}(t)\\
  \end{array}
\right),
\end{equation}
where the diagonal elements of the reduced density operator are given by
\begin{eqnarray}
\rho _{ii}(t)=\left\vert c_{0}\right\vert ^{2}\underset{n=0}{\overset{\infty
}{\sum }}\left\vert \left\langle n\right\vert \varphi _{i}(t)\rangle
\right\vert ^{2}\hspace{0.5cm}(i=0,1,2,3),
\end{eqnarray}
and the off-diagonal elements of the reduced density operator are given by
\begin{eqnarray}
\rho _{ij}(t)&=&\rho^* _{ji}(t) \nonumber\\
&=&c_{i}c^*_{j}\sum^{\infty}_{n=0}\langle n| \varphi _{i}(t)\rangle
\langle \varphi _{j}(t)| n\rangle, \hspace{0.5cm} (i\neq j).
\end{eqnarray}

\section{Micro-micro entanglement witnesss via the geometric phase}

In this section we calculate the geometric phase of the two Rydberg impurities in the BEC when the two impurity qubits are initially in an entangled state while the BEC is in a coherent state. We will obtain a direct relation between the inter-qubit entanglement measured by the concurrence and the geometric phase of the two qubits, and indicate that the geometric phase can witnesss the inter-qubit entanglement.

We assume that two impurity qubits are initially in a Bell-type state while the BEC is initially in a coherent state given by Eq. (8), then the initial state of the impurity-doped BEC is
\begin{eqnarray}
\left\vert \Psi (0)\right\rangle =(\cos \eta _{0}\left\vert 00\right\rangle
+\sin \eta _{0}\left\vert 11\right\rangle )\otimes \left\vert \alpha
\right\rangle.
\end{eqnarray}

Making Eqs. (10) and (13), we can find the reduced density operator of the two qubits in the subspace $\{|00 \rangle, |11\rangle \}$ at an arbitrary time
\begin{eqnarray}
\rho(t)=&\cos ^{2}\eta _{0}\left\vert 00\right\rangle \left\langle
00\right\vert +\sin ^{2}\eta _{0}\left\vert 11\right\rangle \left\langle
11\right\vert \nonumber\\
&+\rho _{01}(t)\left\vert 00\right\rangle \left\langle
11\right\vert +\rho _{10}^{\ast }(t)\left\vert 11\right\rangle \left\langle
00\right\vert,
\end{eqnarray}
where the off-diagonal elements are given by
\begin{eqnarray}
\rho _{01}(t)&=&\rho _{10}^{\ast }(t)\nonumber\\
&=&\frac{1}{2}\sin (2\eta _{0})\overset{%
\infty }{\underset{n=0}{\sum }}\left\langle n\right\vert \varphi
_{0}(t)\rangle \left\langle \varphi _{1}(t)\right\vert n\rangle,
\end{eqnarray}
where $|\varphi _{0}(t)\rangle$ and $|\varphi _{1}(t)\rangle$ are given by Eq. (11).

The off-diagonal element (18)  can be simply expressed as
\begin{eqnarray}
\rho _{01}(t)=\frac{1}{2}\sin (2\eta _{0})e^{i\Lambda _{1}(t)-\Gamma _{1}(t)}.
\end{eqnarray}
where we have introduced the following functions
\begin{eqnarray}
\Lambda _{1}(t)&=&2\omega t+\left\vert \alpha \right\vert ^{2}\sin (2\lambda t), \\
\Gamma _{1}(t)&=&2\left\vert \alpha \right\vert ^{2}\sin ^{2}(\lambda t).
\end{eqnarray}

Then the reduced density operator of the qubits (17) can be rewritten as
\begin{equation}
\rho(t)=\left(
  \begin{array}{cc}
    \cos ^{2}\eta _{0} &\frac{1}{2}\sin (2\eta _{0})e^{i\Lambda _{1}(t)-\Gamma_{1}(t)} \\
    \frac{1}{2}\sin (2\eta _{0})e^{-i\Lambda _{1}(t)-\Gamma _{1}(t)} & \sin^{2}\eta _{0}\\
  \end{array}
\right),
\end{equation}
which has the eigenvalue equation
\begin{eqnarray}
\rho (t)\left\vert \varepsilon _{i}(t)\right\rangle =\varepsilon
_{i}(t)\left\vert \varepsilon _{i}(t)\right\rangle  \hspace{0.5cm}(i=1,2),
\end{eqnarray}
where the eigenvalues are given by
\begin{equation}
\varepsilon _{1,2}=\frac{1}{2}\left[1\pm E(t)\right],
\end{equation}
where $E(t)$ is given by
\begin{align}
E(t)=&\sqrt{1+\sin ^{2}(2\eta _{0})\left[e^{-2\Gamma_1 (t)}-1\right]}.
\end{align}

The corresponding eigenstates are given by
\begin{eqnarray}
\left\vert \varepsilon _{1}(t)\right\rangle&=&\cos \theta (t)\left\vert
00\right\rangle +\sin \theta (t)e^{-i\Lambda (t)}\left\vert
11\right\rangle, \\
\left\vert \varepsilon _{2}(t)\right\rangle&=&\sin \theta (t)\left\vert 00\right\rangle -\cos \theta (t)e^{-i\Lambda (t)}\left\vert
11\right\rangle.
\end{eqnarray}
where the mixing angle in the eigenstates is defined by
\begin{eqnarray}
 \sin\theta (t)&=&\sqrt{\frac{E(t)-\cos (2\eta _{0})}{2E(t)}}, \\
\cos \theta (t)&=&\sqrt{\frac{E(t)+\cos (2\eta _{0})}{2E(t)}}.
\end{eqnarray}

According to the kinematic approach  \cite{Tong},  the geometric phase of two qubits with the reduced density operator of the qubits (22) is given by the following expresion
\begin{eqnarray}
\Phi _{G}&=&\arg \left[\overset{2}{\underset{i=1}{\sum }}\sqrt{\varepsilon
_{i}(0)\varepsilon _{i}(\tau )}\left\langle \varepsilon _{i}(0)\right\vert
\varepsilon _{i}(\tau )\rangle \right. \nonumber\\
&&\times \left. e^{-\int_{0}^{\tau }dt\left\langle
\varepsilon _{i}(t)|\dot{\varepsilon} _{i}(t)\right\rangle }\right],
\end{eqnarray}
where the dot denotes the time derivative, $\tau$ is the  evolution time of the qubits along a quasicyclic path of the impurity qubits, it is determined  by the characteristic frequency of the qubits
\begin{equation}
\tau=\frac{2\pi}{\omega}.
\end{equation}

From Eqs. (21), (24) and (25) we can obtain the initial eigenvales of the the impurity qubits
\begin{align}
\varepsilon _{1}(0)=1, \hspace{0.5cm}\varepsilon _{2}(0)=0.
\end{align}

Then the  geometric phase   (30) becomes
\begin{align}
\Phi _{G}=\arg \left\{\sqrt{\varepsilon _{1}(0)\varepsilon _{1}(\tau )}\langle
\varepsilon _{1}(0)|\varepsilon _{1}(\tau )\rangle e^{-\int_{0}^{\tau
}dt\langle \varepsilon _{1}(t)|\dot{\varepsilon}
_{1}(t)\rangle }\right\},
\end{align}
where the two inner products are given by
\begin{eqnarray}
\langle \varepsilon _{1}(0)|\varepsilon _{1}(\tau )\rangle&=&\cos \eta_{0}\cos \theta (\tau )+e^{-i\Lambda_1 (\tau )}\sin \eta _{0}\sin \theta (\tau), \nonumber\\
\langle \varepsilon _{1}(t)|\dot{\varepsilon}_{1}(t)\rangle
&=&-i\dot{\Lambda}(t)\sin ^{2}\theta (t).
\end{eqnarray}

Substituting Eq. (34) into (33) we can the expression of the  geometric phase
\begin{align}
\Phi _{G}=&\arg\left [\cos \eta _{0}\cos \theta (\tau )+e^{-i\Lambda_1 (\tau )}\sin \eta
_{0}\sin \theta (\tau )\right] \nonumber\\
&+\int_{0}^{\tau }dt\dot{\Lambda_1}(t)\sin ^{2}\theta(t),
\end{align}
where $\Lambda_1 (\tau )$ is given by Eq. (20).

In the following we show the  geometric phase given by Eq. (35) can be used to wintnees the initial entanglement of the impurity qubits. We can use quantum concurrence  to measure the amount of entanglement for an arbitrary quantum state of the two Rydberg impurities. The concurrence of an arbitrary quantum state of two qubits with a density operator $\rho_R(t)$ \cite{33}  is given by
\begin{eqnarray}
\mathcal{C}=\rm{max}\{0,\lambda_1-\lambda_2-\lambda_3-\lambda_4\},
\end{eqnarray}
where the $\lambda_i$ ($i=1,2,3,4$) are the square roots of the eigenvalues in descending order of the operator $R=\rho_R(t)(\sigma_y^1\otimes\sigma_y^2)\rho_R^{\ast}(t)(\sigma_y^1\otimes\sigma_y^2)$ with $\sigma_y$ being the Pauli operator in the computational basis. It ranges from $\mathcal{C}_1=0$ for a separable state to $\mathcal{C}_1=1$ for a maximally entangled state.

If the time-dependent density matrix of a two qubit system can be expressed as
\begin{equation}
\rho(t)=\left(
  \begin{array}{cccc}
    w(t) & 0 & 0 & z(t)\\
    0 & x(t) & 0 & 0\\
    0 & 0 & x(t) & 0\\
    z^{\ast}(t) & 0 & 0 & y(t)\\
  \end{array}
\right)  ,
\end{equation}
one finds that the concurrence corresponding to this state is given by \cite{TYu,34,Wer}
\begin{equation}
\mathcal{C}(t)={\rm max}\{0,2|z(t)|-2x(t)\}.
\end{equation}

For the initial state  (16) of two impurity qubits),  we can obtain the quantum concurrence  with the following expression
\begin{equation}
\mathcal{C}(0)=\big|\sin{(2\eta_0)}\big|.
\end{equation}

Then we can find the relation between the mixing angle and the initial-state entanglement
\begin{eqnarray}
\cos \theta (t)&=&\sqrt{\frac{E(t)+\sqrt{1-\mathcal{C}^{2}}}{E(t)}}, \nonumber\\
 \sin \theta (t)&=&\sqrt{\frac{E(t)-\sqrt{1-\mathcal{C}^{2}}}{E(t)}},
\end{eqnarray}
where $E(t)$ can be expressed as
\begin{eqnarray}
E(t)=\sqrt{1+\mathcal{C}^{2}\left[e^{-2\Gamma _{1}(t)}-1\right]}.
\end{eqnarray}

\begin{figure}[htp]
\centering
\includegraphics[width=8.5cm,height=6.0cm]{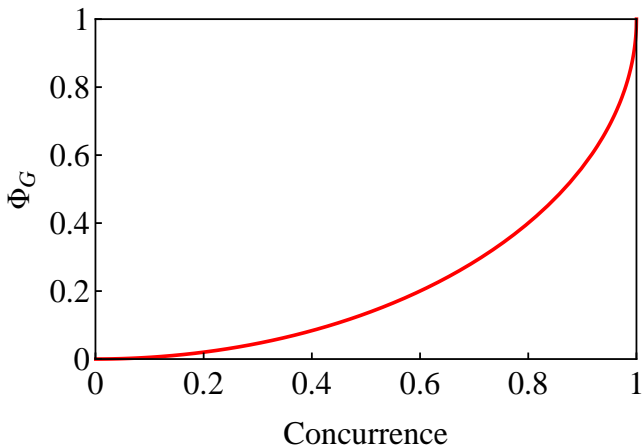}
\caption{(Color online)  The geometric phase of the impurity qubits with respect to the initial inter-qubit entanglement. The geometric phase is scaled by $4\pi\lambda|\alpha|^2/\omega$.} \label{fig4}
\end{figure}

We now consider the case of the qubit-BEC weak coupling regime $\lambda\ll \omega/(2\pi)$, which implies $\lambda\tau \ll 1$.  In the weak coupling regime, we have
\begin{eqnarray}
\Lambda _{1}(t)\approx 2\omega t,  \hspace{0.5cm}  \Gamma_1(t)\approx 0,  \hspace{0.5cm} E(t)\approx 1,
\end{eqnarray}
which leads to
\begin{equation}
\cos \theta (t)=\sqrt{1+\sqrt{1-\mathcal{C}^{2}}}, \hspace{0.5cm} \sin \theta (t)=\sqrt{1-\sqrt{1-\mathcal{C}^{2}}}.
\end{equation}

From Eq. (35) we find that  the geometric phase in the weak coupling regime takes the following simple form
\begin{eqnarray}
\Phi _{G}&=&\frac{4\pi\lambda}{\omega}|\alpha|^2 \left(1-\sqrt{1-\mathcal{C}^{2}}\right),
\end{eqnarray}
which indicates that the geometric phase of the qubits can be manipulated through the coupling strength between the impurity qubits and the BEC or/and the initial-state parameters. In Fig. 2, we have plotted the geometric phase of the impurity qubits with respect to the initial inter-qubit entanglement.  Fig. 2 indicates that the geometric phase of the qubits increases with increasing the initial-state entanglement of the two impurity qubits.

It is interesting to note that we can obtain the direct expression of the initial entanglement of impurity qubts  in terms of the geometric phase
\begin{eqnarray}
 \mathcal{C}&=&1- \left(1 - \frac{\omega\Phi _{G}}{4\pi\lambda |\alpha|^2} \right)^2,
\end{eqnarray}
which indicates that we can see that the accumulated geometric phase of two impurity qubits during the quasicyclic evolution time $\tau=2\pi/\omega$ can witness the initial entanglement of impurity qubts.

\section{Micro-macro entanglement witnesss via the geometric phase}

In this section we investigate the geometric phase of the two  impurity qubits in the BEC when the two impurity qubits are initially  entangled with  the BEC. We will obtain a direct relation between the initial qubit-BEC entanglement measured by the concurrence and the geometric phase of the two qubits, and indicate that the geometric phase can witnesss the hybrid entanglement between the impurity qubits and the BEC.

We assume that two impurity qubits and the BEC are initially in the following hybrid entangled state
\begin{eqnarray}
\left\vert \Psi (0)\right\rangle =\cos \eta _{0}\left\vert 00\right\rangle
\otimes \left\vert \alpha \right\rangle +\sin \eta _{0}\left\vert 11\right\rangle \otimes \left\vert -\alpha \right\rangle \newline,
\end{eqnarray}
where $\left\vert -\alpha\right\rangle $ and $\left\vert \-\alpha\right\rangle $ are two anti-phase coherent states defined by Eq. (8).

At an arbitrary time $t$, the  state of the system under our consideration is a two-component wavefunction
\begin{eqnarray}
\left\vert \Psi (t)\right\rangle &=&\cos \eta _{0}\left\vert 00\right\rangle
\otimes \left\vert \varphi^{'} _{0}(t)\right\rangle +\sin \eta
_{0}\left\vert 11\right\rangle \otimes \left\vert \varphi^{'} _{1}(t)\right\rangle, \nonumber\\
\end{eqnarray}
where the two component wavefunctions of the BEC are given by
\begin{eqnarray}
\left\vert \varphi _{0}^{^{\prime }}(t)\right\rangle &=&e^{-\frac{\left\vert
\alpha \right\vert ^{2}}{2}}\underset{n=0}{\overset{\infty }{\sum }}%
e^{-iE_{00n}t}\frac{\alpha ^{n}}{\sqrt{^{n}!}}\left\vert n\right\rangle,  \\
\left\vert \varphi _{1}^{^{\prime }}(t)\right\rangle&=&e^{-\frac{\left\vert
\alpha \right\vert ^{2}}{2}}\underset{n=0}{\overset{\infty }{\sum }}%
e^{-iE_{11n}t}\frac{(-\alpha )^{n}}{\sqrt{^{n}!}}\left\vert n\right\rangle.
\end{eqnarray}
Above two component wavefunctions of the BEC are generally nonorthogonal with the following inner product
\begin{eqnarray}
\left\langle \varphi^{'} _{0}(t)| \varphi^{'}_{1}(t)\right\rangle&=&e^{-i\left[2\omega t+\left\vert \alpha \right\vert
^{2}\sin (2\lambda t)\right]}e^{\left\vert \alpha \right\vert ^{2}\cos
^{2}(\lambda t)}.
\end{eqnarray}

It is easy to obtain the reduced density operator of the two qubits in the two-state space $\{|00\rangle, |11\rangle \}$ to be
\begin{equation}
\rho(t)=\left(
  \begin{array}{cc}
    \cos ^{2}\eta _{0} & \frac{1}{2}\sin (2\eta _{0})e^{i\Lambda _{2}(t)-\Gamma
_{2}(t)} \\
    \frac{1}{2}\sin (2\eta _{0})e^{-i\Lambda _{2}(t)-\Gamma _{2}(t)} & \sin ^{2}\eta _{0}
  \end{array}
\right),
\end{equation}
where  $\Lambda _{2}(t)$ and $\Gamma _{2}(t)$ are given by
\begin{eqnarray}
\Lambda _{2}(t)&=&2\omega t-\left\vert \alpha \right\vert ^{2}\sin (2\lambda t), \\
\Gamma _{2}(t)&=&2\left\vert \alpha \right\vert ^{2}\cos ^{2}(\lambda t).
\end{eqnarray}

The reduced density operator of the two qubits (51) has the following eigenvalues
\begin{equation}
\tilde{\varepsilon}_{1,2}(t)=\frac{1}{2}\left[1\pm \tilde{E}(t)\right],
\end{equation}
where the function $\tilde{E}(t)$ is defined by
\begin{equation}
\tilde{E}(t)=\sqrt{1+\sin ^{2}(2\eta _{0})\left[e^{-2\Gamma _{2}(t)}-1\right]}.
\end{equation}
The corresponding eigenstates are given by
\begin{eqnarray}
\left\vert \tilde{\varepsilon}_{1}(t)\right\rangle &=&\cos \tilde{\theta}%
(t)\left\vert 00\right\rangle +\sin\tilde{\theta}(t)e^{-i\Lambda
_{2}(t)}\left\vert 11\right\rangle, \\
\left\vert \tilde{\varepsilon}_{2}(t)\right\rangle &=&\sin \tilde{\theta}%
(t)\left\vert 00\right\rangle -\cos \tilde{\theta}(t)e^{-i\Lambda
_{2}(t)}\left\vert 11\right\rangle,
\end{eqnarray}
where the mixing angle functions are defined by
\begin{eqnarray}
\cos \tilde{\theta}(t)&=&\sqrt{\frac{\tilde{E}(t)+\cos (2\eta _{0})}{2\tilde{E} (t)}}, \\
\sin \tilde{\theta}(t)&=&\sqrt{\frac{\tilde{E}(t)-\cos (2\eta _{0})}{2\tilde{E} (t)}}
\end{eqnarray}

Through a calculation similar to the previous section, we can obtain the geometric phase of the two impurity qubits with the following form
\begin{eqnarray}
\tilde{\Phi}_{G}=\tilde{\Phi}_{1}+\tilde{\Phi}_{2},
\end{eqnarray}
where the first part of the geometric phase is given by
\begin{eqnarray}
\tilde{\Phi}_{1}=\arg \left\langle \tilde{\varepsilon}_{1}(0)|\tilde{\varepsilon}_{1}(\tau )\right\rangle  +\int_{0}^{\tau }dt\dot{\Lambda} _{2}(t)\sin ^{2}\tilde{\theta}(t),
\end{eqnarray}
where the inner product $\left\langle \tilde{\varepsilon}_{1}(0)|\tilde{\varepsilon}_{1}(\tau )\right\rangle $ is given by
\begin{eqnarray}
\left\langle \tilde{\varepsilon}_{1}(0)|\tilde{\varepsilon}_{1}(\tau )\right\rangle
&=&  \cos \tilde{\theta}(0)\cos \tilde{\theta}(\tau )+\sin \tilde{\theta}(0)\sin \tilde{\theta}(\tau )\nonumber\\
&&\times e^{i[\Lambda _{2}(0)-\Lambda _{2}(\tau )]}
\end{eqnarray}

The second part of the geometric phase can be expressed as
\begin{eqnarray}
\tilde{\Phi}_{2}=\arg \left\{1+\tilde{F}_{1}(\tau )\tilde{F}%
_{2}(\tau )\tilde{F}_{3}(\tau )\right\}\newline,
\end{eqnarray}
where the three factorization functions are defined by
\begin{eqnarray}
\tilde{F}_{1}(\tau ) &=&\sqrt{\frac{\tilde{\varepsilon}_{2}(0)\tilde{\varepsilon}_{2}(\tau )}{\tilde{\varepsilon}_{1}(0)\tilde{\varepsilon}_{1}(\tau )}}, \\
\tilde{F}_{2}(\tau ) &=&\frac{\langle \tilde{\varepsilon}_{2}(0)|\tilde{\varepsilon}_{2}(\tau )\rangle}{\langle \tilde{\varepsilon}_{1}(0)|\tilde{\varepsilon}_{1}(\tau)\rangle},\\
\tilde{F}_{3}(\tau ) &=&\exp \left\{-\int_{0}^{\tau }dt\left[\langle \tilde{\varepsilon}_{2}(t)| \dot{\tilde{\varepsilon}}_{2}(t)\rangle \right.\right. \nonumber\\
&&\left. \left. -\langle \tilde{\varepsilon}_{1}(t)|\dot{\tilde{\varepsilon}}_{1}(t)\rangle \right]\right\}
\end{eqnarray}

Substituting  Eqs. (54)-(57) into Eqs. (64)-(66), we find that
\begin{eqnarray}
\tilde{F}_{1}(\tau ) &=&\sqrt{\frac{\left[1-\tilde{E}(0)\right]\left[1-\tilde{E}(\tau )\right]}{\left[1+\tilde{E}(0)\right]\left[1+
\tilde{E}(\tau )\right]}}, \\
\tilde{F}_{2}(\tau ) &=&\frac{\cos \tilde{\theta}(0)\cos \tilde{\theta}(\tau )+\sin \tilde{\theta}%
(0)\sin \tilde{\theta}(\tau )e^{i\left[\Lambda _{2}(0)-\Lambda _{2}(\tau )\right]}}{%
\sin \tilde{\theta}(0)\sin \tilde{\theta}(\tau )+\cos \tilde{\theta}(0)\cos
\tilde{\theta}(\tau )e^{i\left[\Lambda _{2}(0)-\Lambda _{2}(\tau )\right]}}, \nonumber\\
\tilde{F}_{3}(\tau ) &=&\exp\left\{i\int_{0}^{\tau }dt\dot{\Lambda}_{2}(t)\cos \left[2\tilde{\theta}(t)\right]\right\}
\end{eqnarray}

In order to observe properties of the geometric phase, we consider the specific case of $\eta_{0}=\frac{\pi}{4}$ and $\lambda\tau=\frac{\pi}{4}$. In this case, we have
\begin{eqnarray}
\Lambda _{2}(0)&=&0, \hspace{0.5cm}\Lambda _{2}(\tau )=2\omega \tau -\left\vert \alpha \right\vert ^{2}, \\
\cos \tilde{\theta}(\tau )&=&\sin \tilde{\theta}(\tau )=\frac{1}{\sqrt{2}},\\
\tilde{E}(0)&=&e^{-2|\alpha|^{2}}, \hspace{0.5cm} \tilde{E}(\tau)=e^{-|\alpha|^{2}},\\
\left\langle \tilde{\varepsilon}_{1}(0)\right\vert \tilde{\varepsilon}_{1}(\tau )\rangle  &=&\cos\left [\frac{1}{2}\Lambda _{2}(\tau )\right]e^{-i\frac{1}{2}\Lambda _{2}(\tau )}
\end{eqnarray}
And the three factorization  functions are given by
\begin{eqnarray}
\tilde{F}_{1}(\tau )&=&\frac{1-e^{-\left\vert \alpha \right\vert ^{2}}}{\sqrt{1+e^{-2\left\vert \alpha \right\vert ^{2}}}},  \\
\tilde{F}_{2}(\tau )&=&\tilde{F}_{3}(\tau )=1.
\end{eqnarray}

Making use of Eqs. (70)-(75), we find that  two parts of the geometric phase of the two qubits are given by
\begin{eqnarray}
\tilde{\Phi}_{1} &=&2\pi \left(\frac{1}{4}+\frac{\lambda }{8}\right)\frac{\left\vert\alpha \right\vert ^{2}}{\pi }, \hspace{0.5cm} \tilde{\Phi}_{2} =0.
\end{eqnarray}
Therefore, the geometric phase of the two qubits takes the following form
\begin{equation}
\tilde{\Phi}_{G}=2\pi \left(\frac{1}{4}+\frac{\omega }{64}\right)\frac{\left\vert\alpha \right\vert ^{2}}{\pi },
\end{equation}
where we have used the condition of $\lambda\tau=\pi/4$.

In order to obtain the direct relation between the geometric phase of the two qubits and the initial entanglement between the two qubits and the BEC, we need to calculate the entanglement amount of the initial state (47).  The initial state given by Eq. (47) is a two-component entangled state. Its entanglement amount can be measured by the quantum concurrence. A calculation similar  to the previous section gives the quantum concurrence of the initial state (47) with the following expression
\begin{equation}
\mathcal{C}(\eta _{0},\alpha )=\left\vert \sin (2\eta _{0})\right\vert\sqrt{1-e^{-2\left\vert \alpha \right\vert ^{2}}}.
\end{equation}

Taking  into account $\eta _{0}=\pi/2$, from Eqs. (76) and (77) we can obtain the direct relation between the geometric phase  and the initial entanglement between the two qubits and the BEC
\begin{equation}
\tilde{\Phi}_{G}=-\frac{1}{64}\left(16+\omega)\ln (1-\mathcal{C} ^{2}\right).
\end{equation}
which indicates that the geometric phase of the qubits can be controlled through changing the initial-state entanglement between the impurity qubits and the BEC. In Fig. 2, we have plotted the geometric phase of the impurity qubits with respect to the initial inter-qubit entanglement.  Fig. 2 indicates that the geometric phase of the qubits increases with increasing the initial-state entanglement  between the impurity qubits and the BEC.

From Eq. (78) we can find that  the initial entanglement between the impurity qubits and the BEC can be directly expressed in terms of the geometric phase as
\begin{equation}
\mathcal{C}=\sqrt{1-\exp\left(-\frac{64}{16+\omega }\tilde{\Phi}_{G}\right)}.
\end{equation}
which indicates that the accumulated geometric phase of two impurity qubits during the quasicyclic evolution time $\tau=2\pi/\omega$ can witness the initial entanglement between the impurity qubits and the BEC.

\begin{figure}[htp]
\centering
\includegraphics[width=8.5cm,height=6.0cm]{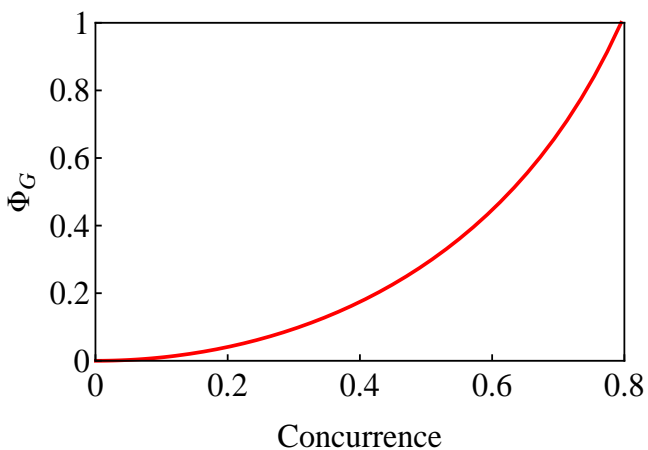}
\caption{(Color online)  The geometric phase of the impurity qubits with respect to the hybrid entanglement between the impurity qubits and the BEC. The geometric phase is scaled by $(16+\omega)/64$.} \label{fig4}
\end{figure}

In what follows we consider another hybrid entangled state between  the impurity qubits and the BEC
\begin{eqnarray}
\left\vert \Psi (0)\right\rangle =\left\vert 0\right\rangle (\cos \eta
_{0}\left\vert 0\right\rangle\otimes \left\vert \alpha \right\rangle +\sin \eta _{0}\left\vert 1\right\rangle \otimes
\left\vert -\alpha \right\rangle ),
\end{eqnarray}
where only the first qubit of the two impurity qubits is entangled with the BEC.

In this case, we   find that at an arbitrary time $t$ the wavefunction of the impurity-doped BEC  is given by
\begin{eqnarray}
\left\vert \Psi (t)\right\rangle &=&\cos \eta _{0}\left\vert 00\right\rangle
\otimes \left\vert \varphi _{0}^{^{\prime }}(t)\right\rangle +\sin \eta
_{0}\left\vert 01\right\rangle \otimes \left\vert \varphi _{01}^{^{\prime}}(t)\right\rangle, \nonumber \\
\end{eqnarray}
where~$\left\vert \varphi _{0}^{^{\prime }}(t)\right\rangle$ is given by Eq.(48) while $\left\vert \varphi _{01}^{^{\prime }}(t)\right\rangle$ is given by
\begin{eqnarray}
\left\vert \varphi _{01}^{^{\prime }}(t)\right\rangle =e^{-\frac{\left\vert
\alpha \right\vert ^{2}}{2}}\underset{n=0}{\overset{\infty }{\sum }}%
e^{-iE_{01n}t}\frac{(-\alpha )^{n}}{\sqrt{n!}}\left\vert n\right\rangle
\newline.
\end{eqnarray}

The inner product the of two component wavefunctions of the BEC is given by
\begin{equation}
\left\langle\varphi'_{0}(t)|\varphi'_{01}(t)\right\rangle =e^{-i\left[(\omega -4J)t+\left\vert \alpha
\right\vert ^{2}\sin (2\lambda t)\right]}e^{-2\left\vert \alpha \right\vert
^{2}\cos ^{2}(\lambda t)}.
\end{equation}

Then we can obtain the reduced density operator of two impurity qubits
\begin{equation}
\rho(t)=\left(
  \begin{array}{cc}
    \cos ^{2}\eta _{0} & \frac{1}{2}\sin (2\eta _{0})e^{i\Lambda _{3}(t)-\Gamma_{3}(t)} \\
    \frac{1}{2}\sin (2\eta _{0})e^{-i\Lambda _{3}(t)-\Gamma _{3}(t)} & \sin ^{2}\eta _{0}
  \end{array}
\right),
\end{equation}
where the phase and the decaying factor are given by
\begin{eqnarray}
\Lambda _{3}(t)&=&(\omega -4J)t-\left\vert \alpha \right\vert ^{2}\sin (2\lambda t), \\
\Gamma _{3}(t)&=&\Gamma _{3}(t)=2\left\vert \alpha \right\vert ^{2}\cos^{2}(\lambda t).
\end{eqnarray}

Two eigenvalues of the reduced density operator (85) are
\begin{eqnarray}
\tilde{\varepsilon}_{1,2}^{^{\prime }}=\frac{1}{2}\left[1\pm\tilde{E}(t)\right],
\end{eqnarray}
where $\tilde{E}(t)$ has been given by Eq.(55).
The corresponding two eigenstates are
\begin{eqnarray}
\left\vert \tilde{\varepsilon}_{1}^{^{\prime }}(t)\right\rangle &=&\cos \tilde{%
\theta}(t)\left\vert 00\right\rangle +\sin \tilde{\theta}(t)e^{-i\Lambda
_{3}(t)}\left\vert 01\right\rangle, \\
\left\vert \tilde{\varepsilon}_{2}^{^{\prime }}(t)\right\rangle&=&\sin \tilde{%
\theta}(t)\left\vert 00\right\rangle -\cos \tilde{\theta}(t)e^{-i\Lambda_{3}(t)}\left\vert 01\right\rangle,
\end{eqnarray}
where the mixing angle functions and $\Lambda_{5}(t)$  are  given by Eqs. (58), (59) and (86), respectively.

We can obtain the geometric phase of the two impurity qubits with the following form
\begin{eqnarray}
\tilde{\Phi}^{'}_{G}=\tilde{\Phi}^{'}_{1}+\tilde{\Phi}^{'}_{2},
\end{eqnarray}
where the first part of the geometric phase is given by
\begin{eqnarray}
\tilde{\Phi}^{'}_{1}=\arg \left\langle \tilde{\varepsilon}^{'}_{1}(0)|\tilde{\varepsilon}^{'}_{1}(\tau )\right\rangle  +\int_{0}^{\tau }dt\dot{\Lambda} _{3}(t)\sin ^{2}\tilde{\theta}(t).
\end{eqnarray}

The second part of the geometric phase can be expressed as
\begin{eqnarray}
\tilde{\Phi}_{2}=\arg \left\{1+\tilde{F}^{'}_{1}(\tau )\tilde{F}^{'}
_{2}(\tau )\tilde{F}^{'}_{3}(\tau )\right\},
\end{eqnarray}
where the three factorization functions are given by
\begin{eqnarray}
\tilde{F}^{'}_{1}(\tau )&=&\tilde{F}(\tau ), \\
\tilde{F}^{'}_{2}(\tau )&=&\frac{\cos \tilde{\theta}(0)\cos \tilde{\theta} (\tau )+\sin \tilde{\theta}(0)\sin \tilde{\theta}(\tau )e^{i\left[\Lambda_{3}(0)-\Lambda _{3}(\tau )\right]}}{\sin \tilde{\theta}(0)\sin \tilde{\theta}(\tau )e+\cos \tilde{\theta}(0)\cos \tilde{\theta}(\tau )e^{i\left[\Lambda_{3}(0)-\Lambda _{3}(\tau )\right]}}, \nonumber \\
&& \\
\tilde{F}^{'}_{3}(\tau )&=&\exp\left \{i\int_{0}^{\tau }dt\tilde{\Lambda}_{3}(t)\cos \left[2\tilde{\theta}(t)\right]\right\}.
\end{eqnarray}

When  we take $\eta_{0}=\frac{\pi}{4}$ and $\lambda\tau=\frac{\pi}{4}$,  we have
\begin{eqnarray}
\tilde{F}_{1}^{^{\prime }}(\tau )&=&\frac{1-e^{-\left\vert \alpha
\right\vert ^{2}}}{\sqrt{1+e^{-2\left\vert \alpha \right\vert ^{2}}}}, \\
\tilde{F}_{2}^{^{\prime }}(\tau )&=&\tilde{F}_{3}^{^{\prime}}(\tau )=1.
\end{eqnarray}

In this case the geometric phase of the two qubits takes the following simple form
\begin{eqnarray}
\tilde{\Phi}_{G}^{^{\prime }}=-\pi \left(1-\frac{4J}{\omega }\right)-\frac{1}{2}%
\left\vert \alpha \right\vert ^{2},
\end{eqnarray}
where we have used $\tau=2\pi/\omega$.

For the hybrid entangled state between the second qubit and the BEC, the quantum concurrence is given by
\begin{equation}
\mathcal{C}(\eta _{0},\alpha )=\left\vert \sin (2\eta _{0})\right\vert
\sqrt{1-e^{-2\left\vert \alpha \right\vert ^{2}}}.
\end{equation}
When $\eta_{0}=\frac{\pi}{4}$ , it becomes
\begin{equation}
\mathcal{C}=\sqrt{1-e^{-2\left\vert \alpha \right\vert ^{2}}}.
\end{equation}

From Eqs. ㄗ99ㄘ and (101) we can obtain the direct relation between the geometric phase and the initial hybrid entanglement
\begin{equation}
\tilde{\Phi}_{G}^{^{\prime }}=-\pi \left(1-\frac{4J}{\omega }\right)+\frac{1}{4}\ln\left(1-\mathcal{C} ^{2}\right).
\end{equation}

Therefore, from Eq. (102) we can express the the initial hybrid entanglement between the two qubits and the BEC in terms of the geometric phase of the two qubits
\begin{equation}
\mathcal{C}=\sqrt{1-\exp \left(4\tilde{\Phi}_{G}^{^{\prime }}-\frac{
16J\pi }{\omega }\right)},
\end{equation}
which indicates that the accumulated geometric phase of two impurity qubits during the quasicyclic evolution time $\tau=2\pi/\omega$ can witness the initial entanglement between the impurity qubits and the BEC for the qubit-BEC system under our consideration.

\section{Concluding remarks}

We have presented a theoretical proposal to witness micro-micro and micro-macro entanglement  in terms of the geometric phase of the impurity qubits in the impurities-doped BEC system, which is a micro-macro quantum system consisting of microscopic impurity qubits and the macroscopic BEC.  We have calculated the geometric phase of the impurity qubits when the two qubits are initially in an entangled state while qubits and the BEC  are initially unentangled, and found that the initial micro-micro entanglement between two impurity qubits can be witnessed in terms of the accumulated geometric phase of two impurity qubits along the quasicyclic  evolution path. When the qubits and the BEC  are initially  in entangled states, i.e., micro-macro entangled states, we have obtained the geometric phase of the qubits for two types of micro-macro entangled states.  We have found a direct relation between the initial micro-macro entanglement and the geometric phase of the impurity qubits, and  demonstrated that the geometric phase can witness micro-micro entanglement between the impurity qubits and the BEC.
The ability of the geometric phase for the impurity qubits to witness micro-micro and micro-macro entanglement in the micro-macro hybrid quantum system provides a new insight for the entanglement detection in hybrid quantum systems which involve micro-macro quantum systems.

\begin{acknowledgements}
{This work is supported by the National Natural
Science Foundation of China under Grants No. 11775075 and No. 11935006, the STI Program of Hunan Province under Grant No. 2020RC4047.}
\end{acknowledgements}



\end{CJK}

\end{document}